\documentclass[aps,prb,floatfix,twocolumn,showpacs]{revtex4}%
\usepackage{amsmath}
\usepackage{graphicx}
\usepackage{amsfonts}
\usepackage{amssymb}%
\providecommand{\U}[1]{\protect\rule{.1in}{.1in}}
\begin{document}
\title{Superconducting/ferromagnetic diffusive bilayer with a spin-active interface:
a numerical study}
\author{Audrey Cottet$^{1}$ and Jacob Linder $^{2}$}
\affiliation{{$^{1}$ Ecole Normale Sup\'{e}rieure, Laboratoire Pierre Aigrain, 24 rue
Lhomond, 75231 Paris Cedex 05, France}}
\affiliation{{$^{2}$ Department of Physics, Norwegian University of Science and Technology,
N-7491 Trondheim, Norway}}


\begin{abstract}
We calculate the density of states (DOS) in a diffusive
superconducting/ferromagnetic bilayer with a spin-active interface. We use a
self-consistent numerical treatment to make a systematic study of the effects
of the Spin-Dependence of Interfacial Phase Shifts (SDIPS) on the
self-consistent superconducting gap and the DOS. Strikingly, we find that the
SDIPS can induce a double gap structure (DGS) in the DOS of the ferromagnet,
even when the superconducing layer is much thicker than the superconducting
coherence lenght. We thus obtain DOS curves which have interesting
similarities with those of Phys. Rev. Lett. \textbf{100}, 237002 (2008).

\end{abstract}
\date{\today}
\maketitle

\section{Introduction}

Superconducting/ferromagnetic ($S/F$) hybrid structures give rise to a
fascinating interplay between two antagonist electronic orders. The
ferromagnetic exchange field $E_{ex}$ privileges one spin direction while
standard superconductivity favors singlet correlations. This leads to a rich
behavior which has triggered an intense activity in the last years (see e.g.
Refs.~\onlinecite{Golubov,BuzdinRMP}). In particular, the ''superconducting
proximity effect'', i.e. the propagation of the superconducting correlations
in ferromagnets, has been widely studied. This propagation is accompanied by
spatial oscillations of the superconducting order parameter, because $E_{ex}$
induces an energy shift between electrons and holes. As a result, one can
build electronic devices with new functionalities, such as Josephson junctions
with negative critical currents\cite{Guichard}, which could find applications
in the field of superconducting circuits\cite{Ioffe,Taro}. From a fundamental
point of view, it is very instructive to study the density of states (DOS) of
$S/F$ structures. So far, this quantity has been less
measured\cite{TakisN,Kontos04,Courtois,Reymond,Sangiorgio} than critical
temperatures or supercurrents. However, this way of probing the
superconducting proximity effect is very interesting because it provides
spectroscopic information. One striking consequence of the spatial
oscillations of the order parameter in the $F$ layer is that the zero-energy
DOS can become larger than in the normal state for certain ferromagnet
thicknesses\cite{TakisN}.

The behavior of $S/F$ hybrid circuits depends crucially on the properties of
the interfaces between the $S$ and $F$ materials. In this paper, we focus on
the case of diffusive structures. Diffusive $S/F$ interfaces have been
initially described with spin-independent boundary conditions\cite{Kuprianov}.
It has been found that the amplitude of the superconducting proximity effect
directly depends on the tunnel conductance $G_{T}$ of an interface (see e.g.
Ref.~\onlinecite{Golubov}). Later, spin-dependent boundary conditions have
been introduced, in the limit of a weakly polarized
ferromagnet\cite{BCHuertas,Cottet:05}. Due to the Spin-Dependence of
Interfacial Phase Shifts (SDIPS)\cite{OtherName,BallisticCase}, one has to
take into account new conductance parameters $G_{\phi}^{F}$ and $G_{\phi}^{S}$
at the $F$ and $S$ sides of the interface, respectively. It has been shown
that $G_{\phi}^{F}$ and $G_{\phi}^{S}$ can significantly affect the behavior
of $S/F$ hybrid circuits. For instance, $G_{\phi}^{F}$ can shift the spatial
oscillations of the superconducting order parameter\cite{Cottet:05}. More
recently, it has been found that $G_{\phi}^{S}$ can induce an effective Zeeman
splitting $\Delta_{Z}^{eff}$ in a superconducting layer with a thickness
$d_{S}$ smaller than the superconducting coherence lengthscale $\xi_{S}%
$\cite{Cottet:07}. This induces a double gap structure (DGS) in the $S$ and
$F$ densities of states. However, in practice, the regime $d_{S}\geq\xi_{S}$
is frequently reached (see e.g. Refs.~\onlinecite{Muhge,Kim}). Remarkably,
DGSs have been recently observed at the $F$ side of \textrm{Ni/Nb} bilayers
with $d_{S}$ much larger than $\xi_{S}$\cite{Sangiorgio}, although
Ref.~\onlinecite{Cottet:07} has found that $\Delta_{Z}^{eff}$ scales with
$d_{S}^{-1}$ in the low $d_{S}$ regime. Whether a DGS persists in the large
$d_{S}$ regime is therefore an important question, especially in the light of
this recent experiment.

In this paper, we study how $G_{\phi}^{F}$ and $G_{\phi}^{S}$ modify the DOS
of a $S/F$ bilayer. We use a numerical treatment to explore a wider parameter
range than in previous works. In particular, we can reach the limit of thick
superconductors and larger values of $G_{\phi}^{S}$. We find that $G_{\phi
}^{S}$ shifts the spatial oscillations of the superconducting order parameter
in $F$, like $G_{\phi}^{F}$. It can also significantly affect the amplitude of
the superconducting gap. When $d_{S}$ increases, the SDIPS-induced DGS becomes
narrower, in agreement with Ref.~\onlinecite{Cottet:07}. Nevertheless, it can
surprisingly persist in the large $d_{S}$ limit. Indeed, on a distance of the
order of $\xi_{S}$ near the $S/F$ interface, the resonance energies of the $S$
spectrum remain spin-dependent because quantum interferences make the
superconducting correlations sensitive to the SDIPS. This behavior is
transmitted to the whole $F$ layer due to the superconducting proximity
effect. We thus obtain, at the $F$ side of $S/F$ bilayers, DOS curves which
have interesting similarities with those of Ref.~\onlinecite
{Sangiorgio}, although $d_{S}\gg\xi_{S}$. More generally, our results could be
useful for interpreting experiments.

This paper is organized as follows. Section II defines the $S/F$ bilayer
problem studied in this article. Section III explains the principle of our
numerical treatment. Section IV presents a detailed study of the SDIPS-induced
DGS. Section V shows the effects of the SDIPS on the self-consistent
superconducting gap and on the oscillations of the zero-energy DOS with the
thickness of $F$. Section VI discusses the data of Ref.~\onlinecite
{Sangiorgio}. Section VII concludes.

\section{Description of the $S/F$ bilayer}

We consider a diffusive $S/F${\LARGE \ }bilayer consisting of a standard
BCS\ superconductor $S$ for $-d_{S}<x<0$, and a ferromagnet $F$ for
$0<x<d_{F}$. We characterize the normal quasiparticle excitations and the
superconducting condensate of pairs with Usadel normal and anomalous Green's
functions $G_{n,\sigma}=\mathrm{sgn}(\omega_{n})\cos(\theta_{n,\sigma})$ and
$F_{n,\sigma}=\sin(\theta_{n,\sigma})$, with $\theta_{n,\sigma}(x)$ the
superconducting pairing angle, which depends on the spin direction $\sigma
\in\{\uparrow,\downarrow\}$, the Matsubara frequency $\omega_{n}(T)=(2n+1)\pi
k_{B}T$, and the spatial coordinate $x$\cite{RevueW}. The Usadel equations
describing the spatial evolution of $\theta_{n,\sigma}$ write%

\begin{equation}
\xi_{S}^{2}\frac{\partial^{2}\theta_{n,\sigma}}{\partial x^{2}}=\frac
{\left\vert \omega_{n}\right\vert }{\Delta_{0}}\sin(\theta_{n,\sigma}%
)-\frac{\Delta(x)}{\Delta_{0}}\cos(\theta_{n,\sigma}) \label{UsadelS}%
\end{equation}
in $S$ and
\begin{equation}
\xi_{F}^{2}\frac{\partial^{2}\theta_{n,\sigma}}{\partial x^{2}}=k_{n,\sigma
}^{2}\sin(\theta_{n,\sigma}) \label{UsadelF2}%
\end{equation}
in $F$, with
\begin{equation}
k_{n,\sigma}=\sqrt{2\left[  i\sigma\mathrm{sgn}(\omega_{n})+\left(  \left\vert
\omega_{n}\right\vert /E_{ex}\right)  \right]  }%
\end{equation}
In the above Eqs., $\Delta_{0}$ denotes the bulk gap of the $S$ material,
$\xi_{S}=(\hbar D_{S}/2\Delta_{BCS})^{1/2}$ the superconducting coherence
lengthscale, $\xi_{F}=(\hbar D_{F}/E_{ex})^{1/2}$ the magnetic coherence
lengthscale, $D_{F(S)}$ the diffusion constant in $F(S)$ and $E_{ex}$ the
ferromagnetic exchange field of $F$. The self-consistent superconducting gap
$\Delta(x)$ occurring in Eq.~(\ref{UsadelS}) can be expressed as
\begin{equation}
\Delta(x)\log[\frac{T}{T_{_{c}}^{0}}]=\frac{\pi k_{B}T}{2}\sum
\limits_{\substack{\sigma\in\{\uparrow,\downarrow\}\\\left\vert \omega
_{n}\right\vert \leq\Omega_{D}}}\left(  \sin(\theta_{n,\sigma})-\frac
{\Delta(x)}{\left\vert \omega_{n}\right\vert }\right)  \label{DeltaBis}%
\end{equation}
with $\Omega_{D}$ the Debye frequency of $S$, $T_{_{c}}^{0}=\Delta_{0}%
\exp(\mathcal{E})/\pi k_{B}$ the bulk transition temperature of $S$, $k_{B}$
the Boltzmann constant, $T$ the temperature and $\mathcal{E}$ the Euler
constant. The above equations must be supplemented with a description of the
boundaries of $S$ and $F$. We use $\left.  \partial\theta_{n,\sigma}/\partial
x\right\vert _{x=-d_{S}^{+}}=\left.  \partial\theta_{n,\sigma}/\partial
x\right\vert _{x=d_{F}^{-}}=0$ for the external sides of the bilayer. For the
$S/F$ interface, we use the spin-dependent boundary
conditions\cite{BCHuertas,Cottet:07}
\begin{equation}
\xi_{F}\left.  \frac{\partial\theta_{n,\sigma}}{\partial x}\right\vert
_{x=0^{+}}=\gamma_{T}\sin[\theta_{n,\sigma}^{F}-\theta_{n,\sigma}^{S}%
]+i\gamma_{\phi}^{F}\sigma\mathrm{sgn}(\omega_{n})\sin[\theta_{n,\sigma}^{F}]
\label{BCright}%
\end{equation}
and
\begin{equation}
\xi_{S}\left.  \frac{\partial\theta_{n,\sigma}}{\partial x}\right\vert
_{x=0^{-}}=\gamma\gamma_{T}\sin[\theta_{n,\sigma}^{F}-\theta_{n,\sigma}%
^{S}]-i\gamma_{\phi}^{S}\sigma\mathrm{sgn}(\omega_{n})\sin[\theta_{n,\sigma
}^{S}] \label{BCleft}%
\end{equation}
with $\theta_{n,\sigma}^{F}=\theta_{n,\sigma}(x=0^{+})$ and $\theta_{n,\sigma
}^{S}=\theta_{n,\sigma}(x=0^{-})$. These equations involve the reduced
conductances $\gamma_{T}=G_{T}\xi_{F}/A\sigma_{F}$ and $\gamma_{\phi}%
^{F(S)}=G_{\phi}^{F(S)}\xi_{F(S)}/A\sigma_{F(S)}$, the barrier asymmetry
coefficient $\gamma=\xi_{S}\sigma_{F}/\xi_{F}\sigma_{S}$, the normal state
conductivity $\sigma_{F(S)}$ of the $F(S)$ material, and the junction area
$A$. Note that we have used a definition of $\gamma_{\phi}^{S}$ which differs
from that of Ref.~\onlinecite{Cottet:07}, to ensure a symmetric treatment of
$\gamma_{\phi}^{F}$ and $\gamma_{\phi}^{S}$ in Eqs.~(\ref{BCright}%
-\ref{BCleft}). The microscopic expressions of the conductances $G_{T}$,
$G_{\phi}^{F}$ and $G_{\phi}^{S}$ can be found e.g. in
Ref.~\onlinecite{Cottet:07}. The term $G_{T}$ corresponds to the usual tunnel
conductance of the interface. The terms $G_{\phi}^{F}$ and $G_{\phi}^{S}$ can
be finite only in case of a Spin-Dependence of Interfacial Phase Shifts
(SDIPS). The SDIPS results from the fact that the scattering phases picked up
by electrons upon scattering by the $S/F$ interface can depend on spin due to
the ferromagnetic exchange field or to a spin-dependent interface potential.
Thus, in principle, any kind of $S/F$ interface can have a finite SDIPS.
However, the exact values of $G_{\phi}^{F}$ and $G_{\phi}^{S}$ are difficult
to predict because they depend on the detailed microscopic structure of the
interface. One possible approach is to consider $G_{\phi}^{F}$ and $G_{\phi
}^{S}$ as fitting parameters which have to be determined from proximity effect
measurements. Note that the derivation of the boundary conditions
(\ref{BCright}-\ref{BCleft}) assumes a weak transmission probability per
channel (tunnel limit), which seems reasonable considering the band structure
mismatch between most $S$ and $F$ materials. It furthermore assumes that the
system is weakly polarized. However, there is no fundamental constraint on the
amplitudes of $G_{T}$, $G_{\phi}^{F}$ and $G_{\phi}^{S}$ because these
parameters consist of a sum of contributions from numerous conducting channels.

In $S/F$ circuits, long-range triplet correlations (between equal spins) can
occur when the circuit includes several $F$ electrodes or domains with
non-colinear magnetizations\cite{bergeret}. Recently, it has been found that
this effect can also arise in $S/F$ circuits with spin-active interfaces, due
to spin-flip interfacial coupling terms which are due e.g. to some misaligned
local moments at the $S/F$ interface\cite{Eschrig}. In our work, we consider
interfaces which are "spin-active" in the sense that the SDIPS is finite.
However, we assume that there is no interfacial spin-flip coupling and that
$F$ is uniformly polarized. Hence, we don't obtain any long-range triplet
component with our model.

\section{Numerical treatment of the problem}

Equations (\ref{UsadelS}-\ref{BCleft}) have already been solved numerically
with a self-consistent procedure in the case $\gamma_{\phi}^{S}=\gamma_{\phi
}^{F}=0$ (see e.g. Ref.~\onlinecite{Gusakova}). In this paper, we study the
case of $\gamma_{\phi}^{S}$ and $\gamma_{\phi}^{F}$ finite, using a numerical
treatment based on a relaxation method. This treatment is divided into two
steps. We first calculate the values of $\Delta(x)$ and $\theta_{n,\sigma}$
self-consistently with a relaxation method in imaginary times. Then, we
determine the pairing angle $\theta_{\sigma}(\varepsilon,x)$ corresponding to
the calculated $\Delta(x)$ by using a similar relaxation method in real times,
i.e. we use $\omega_{n}=-i\varepsilon+\Gamma$ and $\mathrm{sgn}(\omega_{n})=1$
in Eqs.~(\ref{UsadelS}-\ref{BCleft}), with $\varepsilon$ the energy, and
$\Gamma=0.05\Delta_{0}$ a rate which accounts for inelastic processes
\cite{theseW}. Finally we obtain the DOS $N(\varepsilon,x)=\sum
\nolimits_{\sigma}N_{\sigma}(\varepsilon,x)$ at position $x$ by using
$N_{\sigma}(\varepsilon,x)=\left(  N_{0}/2\right)  \operatorname{Re}%
[\cos[\theta_{\sigma}(\varepsilon,x)]]$, with $N_{0}/2$ the normal DOS per
spin direction. Throughout this numerical treatment, we use a discretized
space, with a step of $0.001\xi_{S(F)}$ in $S(F)$. In the following, we mainly
focus on $N_{F}(\varepsilon)=N(\varepsilon,x=d_{F}^{-})$.
Ref.~\onlinecite{Cottet:07} has studied analytically $S/F$ bilayers with
$d_{S}\leq\xi_{S}/2$, $\gamma_{\phi}^{S}\ll1$, $\gamma_{T}\ll1$, and
$d_{F}\geq\xi_{F}$. Our approach allows to go beyond this regime. Note that in
Figures \ref{Fig2} to \ref{Fig1}, the results are shown for $E_{ex}%
=100\Delta_{0}$, $\Omega_{D}=601k_{B}T$ and $k_{B}T=0.1\Delta_{0}$.

\section{The SDIPS-induced Double Gap Structure}

\subsection{Variations of the bilayer spectrum with the thickness of $S$}

\begin{figure}[ptb]
\includegraphics[width=1.\linewidth]{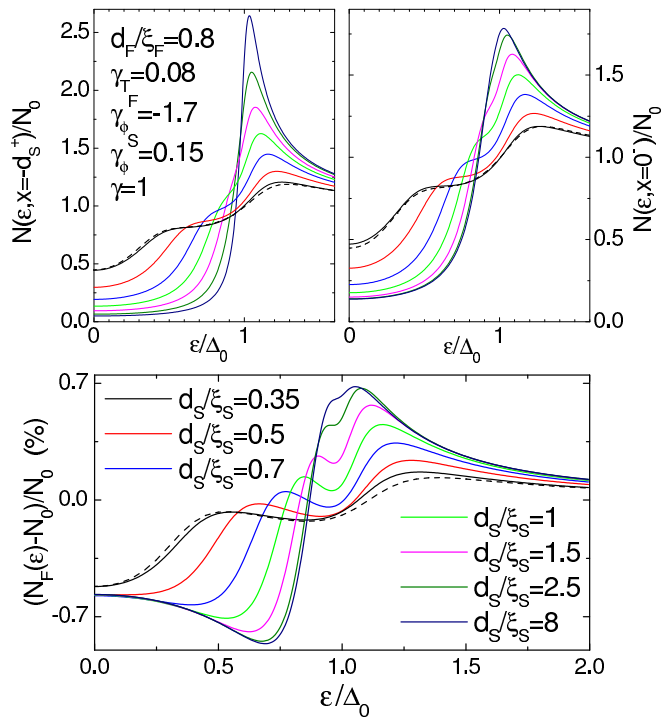}\caption{Densities of states
$N(\varepsilon,x=-d_{S}^{+})$ (top left panel) and $N(\varepsilon,x=0^{-})$
(top right panel) at the left and right side of the superconductor
respectively, and density of states $N_{F}(\varepsilon)$ at the right side of
the ferromagnet (bottom panel), plotted versus $\varepsilon$, for different
values of $d_{S}$. The full lines correspond to our numerical results. The
black dashed lines correspond to the analytical predictions of
Ref.~\onlinecite{Cottet:07}, for $d_{S}/\xi_{S}=0.35$.}%
\label{Fig2}%
\end{figure}

The left and right top panels of Fig.~\ref{Fig2} show the densities of states
$N(\varepsilon,x=-d_{S}^{+})$ and $N(\varepsilon,x=0^{-})$ at the left and
right side of the superconductor respectively, for different values of $d_{S}%
$, and the bottom panel of Fig.~\ref{Fig2} shows the corresponding DOS
$N_{F}(\varepsilon)$ at the right side of $F$. For $d_{S}=0.35\xi_{S}$, the
curves calculated with our numerical code (black full lines) are in close
agreement with the analytical solution given in Ref.~\onlinecite{Cottet:07}
(black dashed lines). The DOS is similar at the two sides of $S$ and displays
a DGS which reveals the existence of an effective Zeeman splitting of the
form
\begin{equation}
\Delta_{Z}^{eff}=2\Delta_{0}\frac{\xi_{S}}{d_{S}}\gamma_{\phi}^{S}=E_{TH}%
^{S}\frac{G_{\phi}^{S}}{G_{S}}\label{heff}%
\end{equation}
with $E_{TH}^{S}=\hbar D_{S}/d_{S}^{2}$ the Thouless energy of the $S$ layer
and $G_{S}=\sigma_{S}A/d_{S}$ its normal state conductance. The DGS is also
visible in $N_{F}(\varepsilon)$ due to the proximity effect. It becomes
narrower when $d_{S}$ increases, in agreement with Eq.~(\ref{heff}) which
indicates that $\Delta_{Z}^{eff}$ scales with $d_{S}^{-1}$. For very large
values of $d_{S}$, the DOS $N(\varepsilon,x=-d_{S}^{+})$ at the left side of
$S$ tends to the bulk\ value $\operatorname{Re}(\cos(\theta_{0}(\varepsilon
)))$, with $\theta_{0}(\varepsilon)=\arctan(\Delta_{0}/(-i\varepsilon
+\Gamma))$ (see top left panel, blue full line). However, a DGS remains
clearly visible in $N_{F}(\varepsilon)$, a result which is quite
counterintuitive considering the low $d_{s}$ expression Eq.~(\ref{heff}) (see
bottom panel, blue full line). Note that in the $S$ layer, with the parameters
of Fig.~\ref{Fig2}, $d_{S}\gg\xi_{S}$ and $\varepsilon=0$ [$\varepsilon
=\Delta_{0}$], $N(\varepsilon,x)$ decays from its bulk value to $N(\varepsilon
,x=0^{-})$ on a distance of the order of $\xi_{S}$ $[2\xi_{S}]$ near the
interface (not shown). In the large $d_{S}$ limit, the DOS $N(\varepsilon
,x=0^{-})$ at the left side of the $S/F$ interface does not show a clear DGS
for the weak value of $\gamma_{\phi}^{S}$ used in Fig.~\ref{Fig2}, because of
the strong DOS\ peak at $\varepsilon=\Delta_{0}$ (see top right panel, blue
full line). However, a DGS would appear more clearly in $N(\varepsilon
,x=0^{-})$ for larger values of $\gamma_{\phi}^{S}$, e.g. using $\gamma_{\phi
}^{S}=0.4$ (not shown). The DGS thus seems to persist at large values of
$d_{S}$ due to an effect which involves a $S$ area with thickness $\sim\xi
_{S}$ near the $S/F$ interface and the whole $F$.

\subsection{Variations the bilayer spectrum with the thickness of $F$}

Due to the ferromagnetic exchange field $E_{ex}$, the zero-energy DOS
$N_{F}(\varepsilon=0)$ oscillates around its normal state value $N_{0}$ when
$d_{F}$ increases\cite{TakisN,Buzdin1,Malek,Bergeret}. In the large $d_{S}$
limit, we find that the DGS can occur at the $F$ side for both an ordinary
($N_{F}(\varepsilon=0)>N_{0}$) and a reversed ($N_{F}(\varepsilon=0)<N_{0}$)
DOS. However, its visibility varies with $d_{F}$, like in the limit $d_{S}%
\leq\xi_{S}/2$ of Ref.~\onlinecite{Cottet:07}. Figure \ref{Fig4}, left panel,
shows $N_{F}$ versus $\varepsilon$, for different values of $d_{F}$ and
$d_{S}/\xi_{S}=5$.\ From $d_{F}=0.53\xi_{F}$ to $0.58\xi_{F}$, $N_{F}%
(\varepsilon=0)-N_{0}$ is positive and the visibility of the DGS increases
(see black, blue and green full lines). For $d_{F}=\xi_{F}$, $N_{F}%
(\varepsilon=0)-N_{0}$ is negative and the DGS is not visible anymore (see red
full line). For $d_{F}=1.6\xi_{F}$, the DGS is visible again, with both the
inner and outer peaks of the DOS inverted due to $N_{F}(\varepsilon
=0)-N_{0}<0$ (see pink dashed line).

\subsection{Variations of the bilayer spectrum with $G_{\phi}^{S}$}

\begin{figure}[ptb]
\includegraphics[width=1.\linewidth]{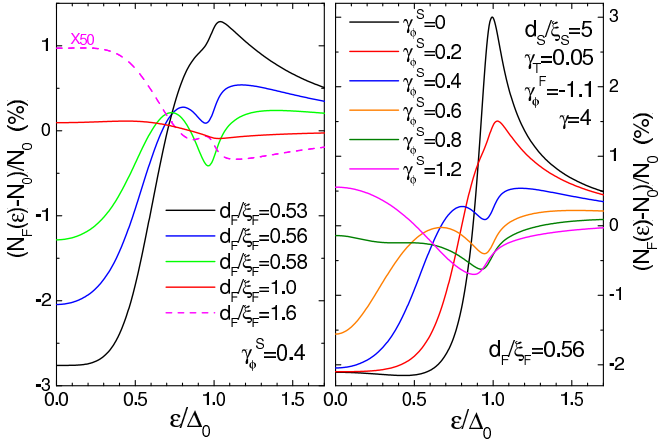}\caption{Density of states
$N_{F}(\varepsilon)$ at the right side of the ferromagnet, plotted versus
$\varepsilon$, for different values of $d_{F}$ (left panel) and different
values of $\gamma_{\phi}^{S}$ (right panel). In the left panel, for $d_{F}%
/\xi_{F}=1.6$, we have multiplied $(N_{F}(\varepsilon)-N_{0})/N_{0}$ by a
factor 50 for visibility of the curve.}%
\label{Fig4}%
\end{figure}

In the parameters range investigated by us (with in particular $E_{ex}%
\gg\Delta$, $0.35\leq d_{S}/\xi_{S}\leq10$ and $0.4\leq d_{F}/\xi_{F}\leq4$),
no DGS occurs when $\gamma_{\phi}^{S}=0$. The DGS studied in this article thus
seems to be a direct consequence of $\gamma_{\phi}^{S}\neq0$ for $d_{S}%
/\xi_{S}$ large as well as $d_{S}/\xi_{S}$ small. Figure \ref{Fig4}, right
panel, shows the variations of $N_{F}(\varepsilon)$ with $\gamma_{\phi}^{S}$,
for a constant value of $d_{F}$. For $\gamma_{\phi}^{S}=0$, no DGS appears.
For a very small $\gamma_{\phi}^{S}$ (see red full line, corresponding to
$\gamma_{\phi}^{S}=0.2$ or $G_{\phi}^{S}=G_{T}$), $N_{F}(\varepsilon)$ shows a
change of slope which corresponds to a smoothed DGS, near $\varepsilon
=\Delta_{0}$ (from the previous section, this DGS occurs only for certain
values of $d_{F}$). When $\gamma_{\phi}^{S}$ becomes sufficiently large,
$N_{F}(\varepsilon)$ shows a clear DGS, i.e. two peaks, one above and one
below $\varepsilon=\Delta_{0}$, while a local minimum is visible for
$\varepsilon\sim\Delta_{0}$. The distance between the two peaks of
$N_{F}(\varepsilon)$ increases with $\gamma_{\phi}^{S}$. In Fig.~\ref{Fig4},
when $\gamma_{\phi}^{S}$ becomes too large ($\gamma_{\phi}^{S}\geq0.8$), the
sign of $N_{F}(\varepsilon=0)-N_{0}$ changes. This suggests that $\gamma
_{\phi}^{S}$ does not only induce DGSs but also shifts the oscillations of
$N_{F}(\varepsilon)$ with $d_{F}$. This last effect will be investigated in
more details for $\varepsilon=0$ in section \ref{Compl}. With the parameters
of Fig.~\ref{Fig4}, the DGS is not visible anymore when $\gamma_{\phi}^{S}$
becomes larger than approximately 1. In the general case, this threshold
strongly depends on the different parameters characterizing the $S/F$ bilayer.

\subsection{Analytic description of the thick superconductor
limit\label{analytic}}

In order to have a better insight on the persistence of the SDIPS-induced DGS
at large values of $d_{S}$, we provide in this section an analytic description
of the case where $S$ is semi-infinite. For simplicity, we assume that the
superconducting gap is only weakly affected by the presence of the $F$ layer,
i.e. $\Delta(x)=\Delta_{0}$. We furthermore assume that the proximity effect
is weak, i.e. $\theta_{n,\sigma}(x\in F)\ll1$ and $\theta_{n,\sigma}(x\in
S)-\theta_{n}^{0}\ll1$, with $\theta_{n}^{0}=\arctan(\Delta_{0}/|\omega_{n}%
|)$. In this case, the Usadel Eqs.~(\ref{UsadelS}-\ref{UsadelF2}) lead to:
\begin{equation}
\theta_{n,\sigma}(x)=\theta_{n,\sigma}^{F}\cosh\left(  \frac{\left[
x-d_{F}\right]  k_{n,\sigma}}{\xi_{F}}\right)  /\cosh\left(  \frac
{d_{F}k_{n,\sigma}}{\xi_{F}}\right)  \label{thetaF}%
\end{equation}
for $x\in F$ and%

\begin{equation}
\theta_{n,\sigma}(x)=\theta_{n}^{0}+\delta\theta_{n,\sigma}^{S}\exp\left(
\frac{x\eta_{n}}{\xi_{S}}\right)  \label{thetaS}%
\end{equation}
for $x\in S$, with $\eta_{n}=\left(  1+(\omega_{n}/\Delta_{0})^{2}\right)
^{1/4}$. We have introduced in the above Eqs. $\delta\theta_{n,\sigma}%
^{S}=\theta_{n,\sigma}(x=0^{-})-\theta_{n}^{0}$ and $\theta_{n,\sigma}%
^{F}=\theta_{n,\sigma}(x=0^{+})$. The linearization of the boundary conditions
(\ref{BCright}-\ref{BCleft}) with respect to these two quantities leads to:%

\begin{equation}
\theta_{n,\sigma}^{F}=\frac{\gamma_{T}\left(  \sin(\theta_{n}^{0})+\cos
(\theta_{n}^{0})\delta\theta_{\sigma}^{S}\right)  }{\gamma_{T}\cos(\theta
_{n}^{0})+i\gamma_{\phi}^{F}\sigma\mathrm{sgn}(\omega_{n})+B_{n,\sigma}}
\label{thetaF2}%
\end{equation}
and%
\begin{equation}
\delta\theta_{n,\sigma}^{S}=-\frac{\gamma\gamma_{T}+i\gamma_{\phi}^{S}%
\sigma\mathrm{sgn}(\omega_{n})}{\eta_{n}^{3}+\left[  \gamma\gamma_{T}%
+i\gamma_{\phi}^{S}\sigma\mathrm{sgn}(\omega_{n})\right]  \frac{\left|
\omega_{n}\right|  }{\Delta_{0}}} \label{deltaThetaS}%
\end{equation}
with $B_{n,\sigma}=k_{n,\sigma}\tanh[d_{F}k_{n,\sigma}/\xi_{F}]$. Importantly,
Eqs.~(\ref{thetaF}-\ref{deltaThetaS}) are valid provided $\delta
\theta_{n,\sigma}^{S}\ll1$ and $\theta_{n,\sigma}^{F}\ll1$, which requires
$\gamma_{T}\ll1$, $\gamma_{\phi}^{S}\ll1$ and $d_{F}\geq\xi_{F}$. We have used
these hypotheses to simplify Eq.~(\ref{deltaThetaS}). The validity of the
approximation $\Delta(x)=\Delta_{0}$ will be discussed in section \ref{Compl}.
\begin{figure}[ptb]
\includegraphics[width=1.0\linewidth]{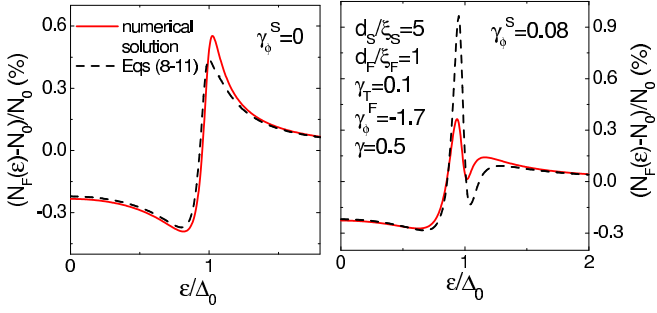}\caption{{}Density of states
$N_{F}(\varepsilon)$ at the right side of the ferromagnet, plotted versus
$\varepsilon$, for $\gamma_{\phi}^{S}=0$ (left panel) and $\gamma_{\phi}^{S}$
finite (right panel). The red lines are calculated from Eqs.~(\ref{thetaF}%
-\ref{deltaThetaS}) and the black dotted lines correspond to the
self-consistent numerical resolution of Eqs.~(\ref{UsadelS}-\ref{BCleft}).}%
\label{Fig5}%
\end{figure}

Figure \ref{Fig5} shows $N_{F}(\varepsilon)$ calculated from the analytic
continuation of Eqs.~(\ref{thetaF}-\ref{deltaThetaS}) (black dashed lines) and
from our numerical code (red full lines), for a weak value of $\gamma$, and
$\gamma_{\phi}^{S}=0$ (left panel) or $\gamma_{\phi}^{S}\neq0$ (right panel).
The two calculations are in relatively good agreement\cite{Fig1and2}. For the
parameters used in Fig.~\ref{Fig5}, we have checked numerically that the
approximation $\Delta(x)=\Delta_{0}$ gives results in very good agreement with
the full resolution of Eqs.~(\ref{UsadelS}-\ref{BCleft}). At $\varepsilon
\sim\Delta_{0}$, small discrepancies arise between the predictions of the
numerical code and of Eqs.~(\ref{thetaF}-\ref{deltaThetaS}), due to resonance
effects which make $\delta\theta_{n,\sigma}^{S}$ and $\theta_{n,\sigma}^{F}$
larger than for $\varepsilon=0$ or $\varepsilon\gg\Delta_{0}$. Equations
(\ref{thetaF}-\ref{deltaThetaS}) allow to recover the fact that a DGS can
appear in $N_{F}(\varepsilon)$, due to $\gamma_{\phi}^{S}\neq0$ (right panel).
In the limit $d_{S}\gg\xi_{S}$, the pairing angle of the system cannot be put
under the form $\theta_{\sigma}(x,\varepsilon)=\theta(x,\varepsilon
-[\sigma\Delta_{Z}^{eff}/2])$, contrarily to what has been found for
$d_{S}\leq\xi_{S}/2$. Therefore, the notion of SDIPS-induced effective Zeeman
splitting is not valid for thick $S$ layers. Nevertheless, from
Eqs.~(\ref{thetaS},\ref{deltaThetaS}), near the S/F interface, the resonance
energies of the $S$ spectrum can be spin-dependent because quantum
interferences make the superconducting correlations sensitive to the SDIPS on
a distance of the order of $\xi_{S}/\eta_{n}$ near the $S/F$ interface. From
Eqs. (\ref{thetaF},\ref{thetaF2}), this behavior is transmitted to the whole
$F$ layer due to the proximity effect. From Eq.~(\ref{deltaThetaS}), the
energy scale related to the occurrence of the DGS has the form:
\begin{equation}
\Delta_{SDIPS}=2\Delta_{0}\frac{G_{\phi}^{S}}{\widetilde{G_{S}}}\label{ThickD}%
\end{equation}
with $\widetilde{G_{S}}=\sigma_{S}A/\xi_{S}$ the normal state conductance of a
slab of thickness $\xi_{S}$ of the $S$ material. Interestingly, this
expression has a form similar to Eq.~(\ref{heff}), with $d_{S}$ replaced by
$\xi_{S}$ (one has $2\Delta_{0}=\hbar D_{S}/\xi_{S}^{2}=\widetilde{E_{TH}^{S}%
}$). Note that for a ballistic $S/F$ single channel contact, the SDIPS also
produces a spin-dependent resonance effect\cite{PRBcottet08}. However, in this
case, one does not obtain a DGS but rather a sub-gap resonance in the
conductance and the zero-frequency noise of the system.

\section{Self-consistent superconducting gap and zero-energy DOS of
$F$\label{Compl}}

\begin{figure}[ptb]
\includegraphics[width=1.0\linewidth]{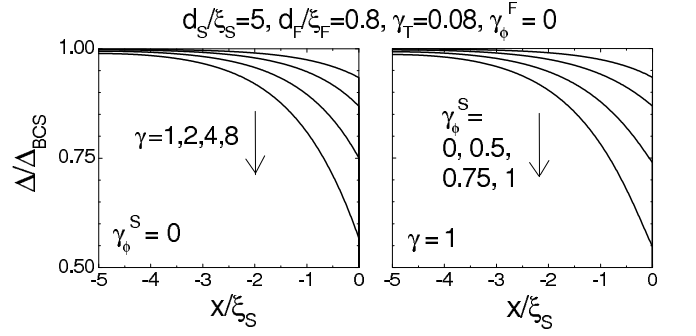}\caption{{}Self-consistent
superconducting gap $\Delta(x)$ versus the spacial coordinate $x$, for
different values of $\gamma$ (left panel) and $\gamma_{\phi}^{S}$ (right
panel).}%
\label{Fig0}%
\end{figure}

For completeness, we now discuss the effects of the SDIPS on the
self-consistent superconducting gap $\Delta(x)$ and the zero-energy DOS
$N_{F}(\varepsilon=0)$ versus $d_{F}$.

It is already known that the amplitude of $\Delta(x)$ decreases when
$\gamma_{T}$ or $\gamma$ increase, similarly to what happens in a $S/$normal
metal bilayer\cite{GolubovSN}. Figure \ref{Fig0} compares the effects of
$\gamma$ (left panel) and $\gamma_{\phi}^{S}$ (right panel) on $\Delta(x)$ (it
only shows the effect of $\gamma_{\phi}^{S}>0$, but the effect of
$\gamma_{\phi}^{S}<0$ is similar). One can see that $\Delta(x)$ significantly
decreases when $|\gamma_{\phi}^{S}|$ increases. Similarly, in a clean
superconductor connected to a ferromagnetic insulator ($FI$), $\Delta(x)$ has
been predicted to decrease due to the spin-dependence of the reflection phases
against $FI$\cite{Tokuyasu}. In contrast, in the regime of parameters
investigated by us, $\gamma_{\phi}^{F}$ has a negligible effect on the value
of $\Delta(x)$ because it does not occur directly in the boundary condition
(\ref{deltaThetaS}) at the $S$ side of the interface. From this brief study of
$\Delta(x)$, we conclude that the approximation $\Delta(x)=\Delta_{0}$ used in
section \ref{analytic} is valid only for sufficiently small values of
$\gamma_{T}$, $\gamma$ and $\gamma_{\phi}^{S}$.

\begin{figure}[ptb]
\includegraphics[width=1.\linewidth]{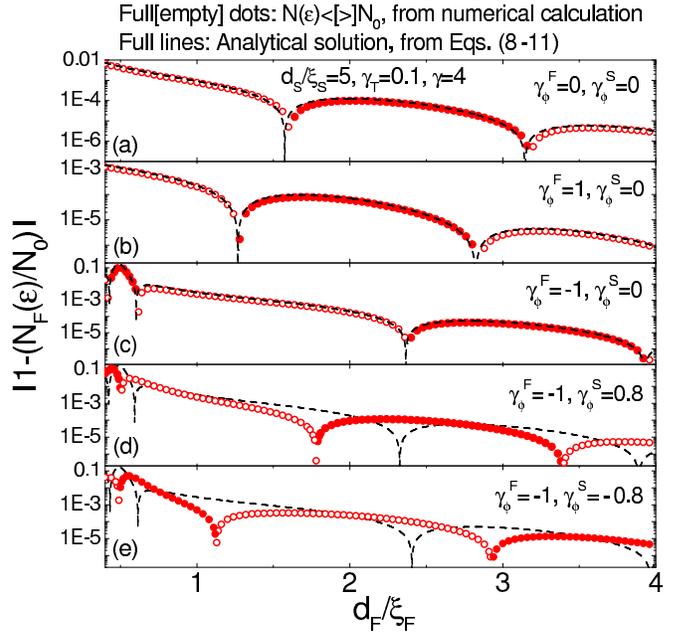}\caption{Zero-energy density
of states $N_{F}(\varepsilon=0)$ at the right side of $F$, versus the
thickness $d_{F}$ of $F$. The density of states calculated numerically is
shown with red dots. The full and empty dots correspond to $N_{F}%
(\varepsilon=0)<0$ and $N_{F}(\varepsilon=0)>0$, respectively. The density of
states given by Eqs.~(\ref{thetaF}-\ref{deltaThetaS}) is shown with black
dashed lines. Panel (a) corresponds to a case with no SDIPS ($\gamma_{\phi
}^{F}=\gamma_{\phi}^{S}=0$). Panels (b) and (c) show the effect of a finite
$\gamma_{\phi}^{F}$. Panels (d) and (e) show the effect of a finite
$\gamma_{\phi}^{S}$, in comparison with panel (c) where $\gamma_{\phi}^{S}=0$.
With the parameters used here, Eqs.~(\ref{thetaF}-\ref{deltaThetaS}) are in
agreement with our numerical code only when $\gamma_{\phi}^{S}=0$.}%
\label{Fig1}%
\end{figure}

Figure \ref{Fig1} presents the effects of $\gamma_{\phi}^{F}$ and
$\gamma_{\phi}^{S}$ on the variations of $N_{F}(\varepsilon=0)$ with $d_{F}$.
The DOS calculated numerically is shown with symbols and the DOS given by
Eqs.~(\ref{thetaF}-\ref{deltaThetaS}) is shown with full lines. In panels (a),
(b) and (c), we have used $\gamma_{\phi}^{S}=0$, so that the two calculations
are in close agreement. In panels (d) and (e), the two calculations strongly
differ because $\gamma_{\phi}^{S}$ is too large for the hypotheses leading to
Eqs.~(\ref{thetaF}-\ref{deltaThetaS}) to be valid. We recover the fact that,
in the regime $d_{F}\geq\xi_{F}$, $N_{F}(\varepsilon=0)$ shows exponentially
damped oscillations with $d_{F}$ \cite{TakisN,Buzdin1}. In the regime
$d_{F}\leq\xi_{F}$, the oscillations of $N_{F}(\varepsilon=0)$ with $d_{F}$
are less regular. This can be understood from the analytic description of
Section \ref{analytic}. For $d_{F}\geq\xi_{F}$, one has $B_{n,\sigma}\sim
k_{n,\sigma}$, so that $N_{F}(\varepsilon=0)$ depends on $d_{F}$ through the
$\cosh\left(  d_{F}k_{n,\sigma}/\xi_{F}\right)  $ term of Eq.~(\ref{thetaF})
only. For $d_{F}\leq\xi_{F}$, $B_{n,\sigma}$ and thus $\theta_{n,\sigma}^{F}$
strongly depend on $d_{F}$, which complicates the variations of $N_{F}%
(\varepsilon=0)$ with $d_{F}$ and leads to more irregular oscillations.
Ref.~\onlinecite{Cottet:05} has already shown that $\gamma_{\phi}^{F}$ can
shift the oscillations of $N_{F}(\varepsilon=0)$ with $d_{F}$. Panels (b) and
(c) confirm this result and also shows that a positive (negative)
$\gamma_{\phi}^{F}$ decreases (increases) the amplitude of $N_{F}%
(\varepsilon=0)$. From panels (d) and (e), $\gamma_{\phi}^{S}$ can also
significantly shift the oscillations of $N_{F}(\varepsilon=0)$ with $d_{F}$,
in agreement with Fig.~\ref{Fig4}, right panel. For the parameters used in
Fig.~\ref{Fig1}, $\gamma_{\phi}^{S}$ does not modify spectacularly the
amplitude of $N_{F}(\varepsilon)$. For larger values of $|\gamma_{\phi}^{S}|$,
the amplitude of the superconducting proximity effect would significantly
decrease due to a reduction of $\Delta(x)$ (not shown).

\section{Discussion on the data of Phys. Rev. Lett. \textbf{100}, 237002
(2008)}

We now consider the DOS measurements realized by SanGiorgio et al. for
\textrm{Nb/Ni} bilayers with $d_{S}=50$~\textrm{nm}\cite{Sangiorgio}. From
Ref.~\onlinecite{Kim} which considers samples fabricated by the same team, one
has $\xi_{S}\sim10$~\textrm{nm}, so that $d_{S}/\xi_{S}\sim5$. However, double
gap structures have been observed by SanGiorgio et al., which motivates a
comparison with our model.

\begin{figure}[ptb]
\includegraphics[width=0.95\linewidth]{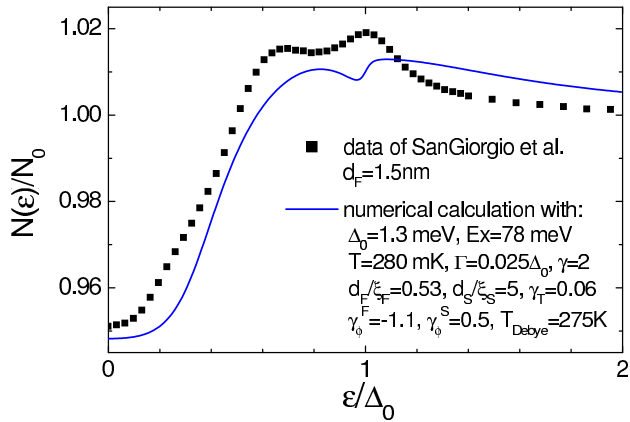}\caption{{}Comparison
between the data of SanGiorgio et al., Phys. Rev. Lett. \textbf{100}, 237002
(2008), for $d_{s}=1.5$~\textrm{nm}, and our numerical calculation with
$\Delta_{0}=1.3$~\textrm{meV}, $E_{ex}=78$~\textrm{meV}, $k_{B}T=280$%
~\textrm{mK}, $d_{F}/\xi_{F}=0.53$, $d_{S}/\xi_{S}=5$, $\gamma_{T}=0.06$,
$\gamma_{\phi}^{F}=-1.1$, $\gamma_{\phi}^{S}=0.5$, $\gamma=2$, $T_{Debye}%
=275$~\textrm{K} and $\Gamma=0.025\Delta_{0}$.}%
\label{Fig6}%
\end{figure}

Figure~\ref{Fig6} compares the data measured for $d_{F}=1.5$~\textrm{nm}
(black squares) with our numerical calculation (blue full line). Our
calculation reproduces almost quantitatively the experimental curve. We have
used $d_{F}=1.5$~\textrm{nm, }$d_{S}/\xi_{S}=5$ and $T=280$~\textrm{mK}, in
agreement with Refs.~\onlinecite{Sangiorgio,Kim}. We have also used the
exchange field $E_{ex}=78$~\textrm{meV}, estimated by
Ref.~\onlinecite{Sangiorgio}, and the Debye temperature $T_{D}=275$~\textrm{K}
of \textrm{Nb,} taken from Ref.~\onlinecite{Ashcroft}. We have assumed
$\xi_{F}=2.83$~\textrm{nm}, $\gamma_{T}=0.06$, $\gamma_{\phi}^{F}=-1.1$,
$\gamma_{\phi}^{S}=0.5$, and $\Gamma=0.025\Delta_{0}$. Note that from
Ref.~\onlinecite{Kim}, one has $\sigma_{S}^{-1}\sim15.9$ $\mathrm{\mu\Omega}%
.$\textrm{cm} and $\sigma_{F}^{-1}\sim9.7$ $\mathrm{\mu\Omega}.$\textrm{cm},
so that one should have $\gamma=\xi_{S}\sigma_{F}/\xi_{F}\sigma_{S}\sim2.9$
with the above values of $\xi_{S}$ and $\xi_{F}$. In Figure~\ref{Fig6}, we
have used a value $\gamma=2$, which is in relatively good agreement with this
estimate. The values of $\gamma_{T}$, $\gamma_{\phi}^{F}$, $\gamma_{\phi}^{S}$
and $\gamma$ used in the fit yield $G_{\phi}^{F}/G_{T}\sim18$ and $G_{\phi
}^{S}/G_{T}\sim4$. A theoretical prediction of these ratios is very difficult
because they depend on the detailed microscopic structure of the
\textrm{Nb/Ni} interface. However, with a simple delta function barrier model,
it already is possible to find situations where $G_{\phi}^{S}$ and $G_{\phi
}^{F}$ are larger than $G_{T}$ (see Appendix A and Fig. 6 of
Ref.~\onlinecite{Cottet:07}). Therefore, we think that the SDIPS parameters
used by us are possible.

We cannot reproduce quantitatively the data obtained by SanGiorgio et al. for
all values of $d_{F}$ with the same set of parameters as for $d_{F}%
=1.5$~\textrm{nm}. We think that this might be due to the fact that some
characteristics of the samples like e.g. $E_{ex}$ and thus $\xi_{F}$,
$\gamma_{\phi}^{F}$, $\gamma_{\phi}^{S}$, and $\gamma$ can vary with $d_{F}%
$\cite{thesisTK}. In the data of SanGiorgio et al., from $d_{F}=1.5$%
~\textrm{nm} to $d_{F}=3.0$~\textrm{nm}, the distance between the two peaks of
the DGS increases, like in our model (see Fig.~\ref{Fig4}, left panel).
However, the outer peak of the DGS remains very close to $\varepsilon
\sim\Delta_{0}$, which seems difficult to reproduce with our model. Note that
in Ref.~\onlinecite{Sangiorgio}, for $d_{F}=3.5$~\textrm{nm}, the outer peaks
of the DOS are inverted, whereas a sharp dip occurs at low energies, which can
give the impression that the innner peaks of the DOS persist but are not
inverted, in contrast to what we find (see section IV.B). However, we think
that the observation of this zero-bias dip is not totally reliable. Indeed,
SanGiorgio et al. explain that, in the DOS of their thickest samples (for
$d_{F}>3.5$~\textrm{nm}), "the zero bias peak is due to the steep voltage
dependence of the background conductance and is therefore a by-product of the
data normalization procedure". The zero bias-dip of the $d_{F}=3.5$%
~\textrm{nm} sample occurs on the same energy scale as these zero-bias peaks,
and it goes together with a strange zero-bias singularity similar to those
shown by the thickest samples. Therefore, we are not sure whether a proper
interpretation of the data of Ref.~\onlinecite{Sangiorgio} must take into
account this feature. Ref.~\onlinecite{VolkovEfetov} suggests that the DGS
observed by Sangiorgio could be due to triplet correlations. However, it is
difficult to know whether this interpretation can be more satisfying than ours
because Ref.~\onlinecite{VolkovEfetov} does not shown any quantitative
interpretation of the data and does not discuss, for instance, the evolution
of the inner and outer peak positions with $d_{F}$.

\section{Conclusion}

In summary, we have calculated the density of states (DOS) in a diffusive
superconducting/ferromagnetic ($S/F$) bilayer with a spin-active interface. We
have used a self-consistent numerical treatment to make a systematic study of
the effects of the Spin-Dependence of Interfacial Phase Shifts (SDIPS). We
characterize the SDIPS with two conductance-like parameters $G_{\phi}^{S}$ and
$G_{\phi}^{F}$, which occur in the boundary conditions describing the $S$ and
$F$ sides of the interface, respectively. We find that the amplitude of
$\Delta(x)$ significantly decreases if $G_{\phi}^{S}$ is too strong, whereas
it is almost insensitive to $G_{\phi}^{F}$. In contrast, both $G_{\phi}^{S}$
and $G_{\phi}^{F}$ can shift the oscillations of the zero-energy DOS of $F$
with the thickness of $F$. Remarkably, we find that the SDIPS can produce a
double gap structure in the DOS of $F$, even when the $S$ layer is much
thicker than the superconducting coherence lenght. This leads to DOS curves
which have striking similarities with those of Ref.~\onlinecite{Cottet:05}.
More generally, our results could be useful for interpreting future
experiments on superconducting/ferromagnetic diffusive hybrid structures.

We acknowledge interesting discussions with A. A. Golubov, T. Kontos, M. Yu.
Kuprianov and N. Regnault. We thank P. SanGiorgio and M. Beasley for
discussions and for sending us their data. J.L. was supported by the Research
Council of Norway, Grants No.158518/431 and No. 158547/431 (NANOMAT), and
Grant No. 167498/V30 (STORFORSK).


\begin{thebibliography}{99}
\bibitem {OtherName}The SDIPS is called by some other authors
\textquotedblright spin-mixing angle\textquotedblright\ or \textquotedblright
spin-rotation angle\textquotedblright. However, in our case, we consider a
single ferromagnet which is uniformly polarized, so that the notions of mixing
or rotation are not relevant.
\bibitem {BallisticCase}The effects of the SDIPS on $S/F$ systems have been
initially studied in the ballistic case, see e.g. Refs.~\onlinecite
{Millis,Tokuyasu}.
\bibitem {Fig1and2}For the parameters of Figs. \ref{Fig2} and \ref{Fig4},
Equations (\ref{thetaF}-\ref{deltaThetaS}) are not in good agreement with our
numerical results because the hypotheses leading to these equations are not
satisfied (see section \ref{Compl}.)
\end{thebibliography}
\end{document}